\documentclass[%
 reprint,
superscriptaddress,
showpacs,
nofootinbib, nobibnotes, 
 amsmath,
 amssymb,
 aps,
 prl
]{revtex4-1}

\usepackage{blindtext}
\usepackage{dcolumn}
\usepackage{bm}
\usepackage{color, soul, colortbl} 
\usepackage[colorlinks=true,
						linkcolor=blue,
						urlcolor=blue,
						citecolor=blue,
						bookmarks=true,
						pdfborder={0 0 0}]{hyperref}
       
\usepackage[rgb]{xcolor}  

\usepackage{graphicx,epsfig}
\usepackage{subfigure}
\usepackage{amsmath}
\usepackage{xfrac}

\begin{document}

\preprint{APS/123-QED}

\title{Network Dynamics of Innovation Processes}

\author{Iacopo Iacopini}
    \affiliation{School of Mathematical Sciences, Queen Mary University of London, London E1 4NS, United Kingdom}
	\affiliation{The Alan Turing Institute, The British Library, London NW1 2DB, United Kingdom}
\author{Sta\v{s}a Milojevi\'{c}}
	\affiliation{Center for Complex Networks and Systems Research, School of Informatics and Computing, Indiana University, Bloomington, Indiana 47408, USA}
\author{Vito Latora}
\affiliation{School of Mathematical Sciences, Queen Mary University of London, London E1 4NS, United Kingdom}
\affiliation{Dipartimento di Fisica ed Astronomia, Universit\`a di Catania and INFN, I-95123 Catania, Italy}

\date{\today}

\begin{abstract}

  We introduce a model for the emergence of innovations, in which cognitive processes are described
  as random walks on the network of links among ideas or concepts, and an innovation corresponds to the 
first visit of a node. 
The transition matrix of the random walk depends on the network weights,
while in turn the weight of an edge is reinforced by the passage of a
walker. The presence of the network naturally accounts for the
mechanism of the ``adjacent possible," and the model reproduces both
the rate at which novelties emerge and the correlations among them
observed empirically. We show this by using synthetic networks and by
studying real data sets on the growth of knowledge in different
scientific disciplines. Edge-reinforced random walks on complex
topologies offer a new modeling framework for the dynamics of
correlated novelties and are another example of coevolution of
processes and networks.
\end{abstract}

\maketitle

Creativity and innovation are the underlying forces driving the growth
of our society and economy. Studying creative processes and
understanding how new ideas emerge and how novelties can
trigger further discoveries is therefore fundamental if we want to
devise effective interventions to nurture the success and sustainable
growth of our society.
Recent empirical studies have investigated the emergence of
novelties in a wide variety of different contexts, including 
science \cite{rzhetsky2015choosing,sinatra2016quantifying}, 
knowledge and information \cite{andjelkovic2016topology,rodi2017search},  
goods and products \cite{saracco2015innovation}, language \cite{puglisi2008cultural}, and 
also gastronomy \cite{fink2016dynamics} and cinema \cite{sreenivasan2013quantitative}.
In particular, the authors of Refs.~\cite{tria2014dynamics, monechi2017waves,
	cattuto2007semiotic, cattuto2007vocabulary} have looked at different
types of temporally ordered sequences of data, such as sequences of words,
songs, Wikipages and tags to study how the number $S(t)$ of novelties grows
with the length of the sequence $t$.
They have found that the Heaps' law, i.e. a power-law behaviour $S(t)
\sim t ^{\beta}$ originally introduced to describe the number of
distinct words in a text document \cite{heaps1978information}, applies
to different contexts, producing different values of $\beta<1$.
In parallel to the empirical analyses, various models have
been proposed to reproduce the innovation dynamics in different 
domains, such as linguistics \cite{gerlach2013stochastic, lu2013deviation},
social systems \cite{dankulov2015dynamics}, or self-organized
criticality (SOC) \cite{tadic2017mechanisms}. Other approaches have modeled
the emergence of innovation as an evolutionary process, such as the
Schumpeterian economic dynamics proposed by Thurner {\em et al.}
\cite{thurner2010schumpeterian} and the evolutionary game among
innovators and developers proposed by Armano and Javarone
\cite{armano2017beneficial}. 
Urn models are another useful framework to 
study innovation processes in evolutionary biology,
chemistry, sociology, economy and text analysis \cite{simkin2011re,
	marengo2016arrival}. In the classic Polya urn model
\cite{hoppe1984polya,polya1930quelques}, a temporal sequence of
discoveries can be generated by drawing balls from an urn that contains
all possible inventions. Several variations
have been proposed, such as the urn model with memory, to reproduce the
dynamics of collaborative tagging \cite{cattuto2007semiotic}, or the
more recent model by Tria and co-workers \cite{tria2014dynamics, loreto2016dynamics}, 
which adds the concept of the adjacent possible \cite{kauffman1996investigations,
		gravino2016crossing} to the reinforcement mechanism of the Polya's urn framework.

In this letter, we propose to model the dynamical mechanisms leading to discoveries and innovations
as an edge-reinforced random walk (ERRW) on an underlying network of relations among concepts and ideas.
Random walks on complex networks
\cite{albert2002statistical,newman2003structure,boccaletti2006complex,barrat2008dynamical,latora_nicosia_russo_2017}
have been studied at length \cite{masuda2016random}.  In the context
of innovation, they have been used to build exploration models for
social annotation \cite{cattuto2009collective}, music album
popularity \cite{Monechi170433}, knowledge acquisition
\cite{de2017knowledge}, human language complexity \cite{allegrini2004intermittency} and evolution in research interests
\cite{jia2017quantifying}. A special class of random walks are those
with reinforcement \cite{gomez2008entropy, agliari2012true, pemantle2007survey},
which have been successfully applied to biology \cite{boyer2014random}
and mobility \cite{choi2012modeling, szell2012understanding}. 
In particular, the concept of edge reinforcement 
\cite{merkl2006linearly,keane2000edge,foster2009reinforced} was introduced in the
mathematical literature by Coppersmith and Diaconis
\cite{Coppersmith1952}. Here, we will use ERWWs 
to mimic how different concepts are
explored moving from a concept to an adjacent one in the network, with
innovations being represented, in this framework, by the first
discovery of nodes. As supported by empirical
observations, we expect indeed the walkers to move more frequently among already known
concepts and, from time to time, to discover
new nodes. For this reason, we introduce and study a model in which
the network is co-evolving with the dynamical
process taking place over it. In our model,    
(i) random walkers move over a network with assigned topology and
whose edge weights represent the strength of concept associations, and  
(ii) the network evolves in time through a
reinforcement mechanism in which the weight of an edge is increased
every time the edge is traversed by a walker, making traversed edges more likely to be traversed again.
As we will show, this model is able to reproduce the statistical properties observed in real data of innovation processes, i. e., the Heaps' law \cite{heaps1978information}, and by tuning the amount of reinforcement it can give rise to different scaling exponents.  
Furthermore, correlations in the temporal sequences     
of visited concepts and innovations will appear as
a natural consequence of the interplay between the network
topology and the reinforcement mechanism that controls
the exploration dynamics.
\begin{figure}
	\vspace{-1em}
	\centering
	\includegraphics[width=0.48\textwidth]
	{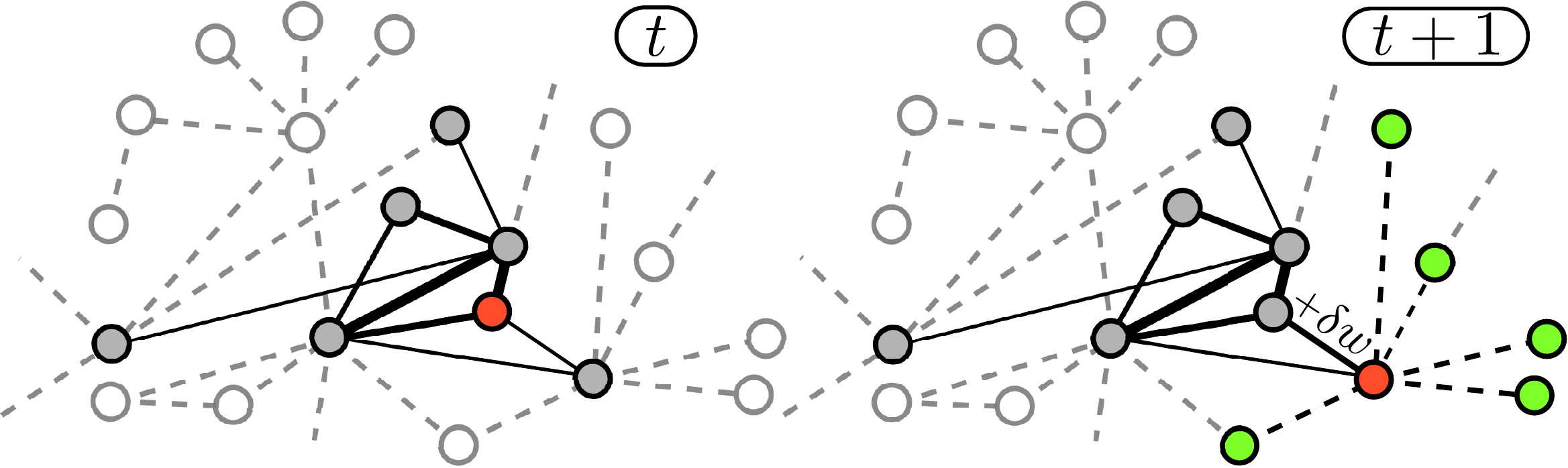}
	\vspace{-1em}
	\caption{\label{fig:network_illustration} Edge-reinforced random walks (ERRWs)
		produce a coevolution of the network with the dynamics of the
		walkers. At time $t$ the walker is on the red node
		and has already visited the gray nodes, while the shaded nodes are
		still unexplored. The widths of edges are
		proportional to their weights. At time $t+1$ the
		walker has moved to a neighbor (red) with probability
		as in Eq.~(\ref{eq:jumping}), and the weight of the used edge has
		been reinforced by $\delta w$. At this point, the walker
		will preferentially go back, although it can also access the set of ``adjacent possible'' (green).}
	\vspace{-1em}
\end{figure}

\medskip
\textit{Model.} Let us consider a random walker over a weighted
connected graph $G(\mathcal{V},\mathcal{E})$, where $\mathcal{V}$ and $\mathcal{E}$ are,
respectively, a set of $N=|\mathcal{V}|$ nodes and a set of $K=|\mathcal{E}|$ links.
Each node of the graph represents a concept or an idea, and the presence of 
a link $(i,j)$ denotes the existence of a direct relation 
between two concepts $i$ and $j$. The values of $N$ and $K$ and the topology of the network
are assumed to be
fixed, while the weights of the edges can 
change in time according to the dynamics of the walker, which, as we will see below, is in turn  influenced by the underlying network. 
The graph at time $t$, with $t=0,1,2,\ldots$, is fully described by the non-negative time-dependent adjacency matrix $W^{t} \equiv \{w^{t}_{ij}\}$, where the value $w^{t}_{ij}$ is different 
from 0 if the two concepts $i$ and $j$ are related, and quantifies the strength of the relationship at time $t$. We initialize the network assuming that at time $t=0$ all the edges have the same weight, namely $w^{0}_{ij}=1 ~ \forall (i,j)\in \mathcal{E}$. 
The dynamics of the walkers is defined as follows: at each time step $t$, a walker at node $i$ jumps to a randomly chosen neighboring node $j$ with a probability
proportional to the weight of the connecting edge. 
Formally, the probability of going from node $i$ to node $j$ at time $t$ is:  
\begin{equation}\label{eq:jumping}
\text{Prob}^{t}(i\rightarrow j)=\pi^{t}_{ji}=\frac{w^{t}_{ij}}{\sum_{l}w^{t}_{il}}
\end{equation}
where the time-dependent transition probability matrix 
$\Pi^{t}\equiv  \{ \pi^{t}_{ij} \}$ depends on the 
weights of all links
at time $t$ \cite{cover2012elements}. 
The transition probabilities satisfy the normalization $\sum_{j} \pi^t_{ji}=1$ $\forall i, t$, 
and we assume that $G$ has no self-loops, so that the walker changes 
position at each time step. On the other hand, 
the network coevolves with the random 
	walk process, since  every time a walker traverses a link, it increases its weight by a quantity 
$\delta w>0$, as illustrated in Fig.~\ref{fig:network_illustration}.  
This mechanism mimics the fact that the relation between two concepts is 
reinforced every time the two concepts are associated by a cognitive 
process.
Formally, the dynamics of the network is the following. 
Every time an edge $(i,j) \in\mathcal{E}$ is traversed at time $t$, the associated weight is reinforced as 
\begin{equation}\label{eq:reinforcement}
w^{t+1}_{ij}=w^{t}_{ij}+\delta w
\end{equation}
The quantity $\delta w$, called reinforcement, is the only tunable parameter of the model. The idea of a walker preferentially returning on its steps is in line with the classical rich-get-richer 
paradigm, which has been extensively used in the network literature to grow scale-free graphs 
\cite{barabasi1999emergence}, and is here implemented in terms of reinforcement of the edges, instead of using a random
walk biased on some properties of the nodes \cite{gomez2008entropy, bonaventura2014characteristic, sood2007localization}.

%
%
\noindent The coevolution of network and walker motion induces
a long-term memory in the trajectories which reproduces, as we will
show below, the empirically observed correlations in the dynamics of
innovations \cite{tria2014dynamics}. In fact, if $i_{t}$ is a
realization of the random variable $X_t$ denoting the position of the
walker at time $t$, the conditional probability
$\text{Prob}\left[X_{t+1}=i|i_{0},i_{1},\ldots,i_{t}\right] $ that, at
time step $t+1$, the walker is at node $i$, after a trajectory
$\mathcal{S}=(i_{0},i_{1},i_{2},\ldots,i_t)$, depends on the whole
history of the visited nodes, namely on the frequency but also on the
precise order in which they have been visited
\cite{szell2012understanding}. The strongly non-Markovian
\cite{gardiner1985stochastic} nature of the random walks comes indeed
from the fact that the transition matrix $\Pi^{t}$ coevolves with the rearrangement of the weights. 
This makes our
approach fundamentally different from the other models based on Polya-like
processes.
For instance, in the Tria {\em et al.}~urn model
\cite{tria2014dynamics}, where an innovation corresponds to the
extraction of a ball of a new color, the probability of extracting a
	given color (colors correspond to node labels in our model) at time
	$t+1$ only depends on the number of times each color has been extracted
	up to time $t$, and not on the precise sequence of colors.
Moreover, the use of an underlying network (see
Fig.~\ref{fig:network_illustration}) is a natural way to include the
concept of the {\em adjacent possible} in our model, without the need
of a triggering mechanism and further parameters, which are instead
necessary in the UM (balls of new colors added into the urn whenever a
color is drawn out for the first time) and in its mapping in terms of growing graphs considered in SI of Refs.~\cite{tria2014dynamics,monechi2017waves}. 

\begin{figure}[t]
	\vspace{-1em}
	\centering
	\includegraphics[width=0.48\textwidth]
	{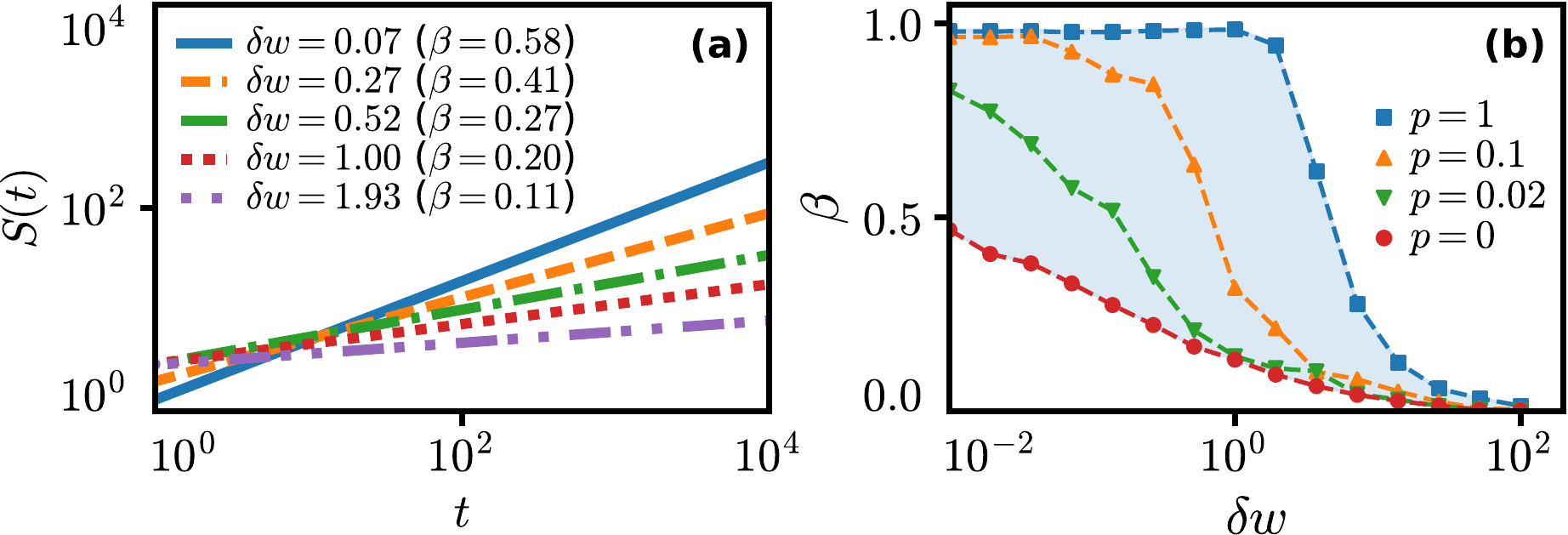}
	\vspace{-1em}
	\caption{\label{fig:sw_model} ERRW on SW networks with $N=10^5$ and $m=1$. \textbf{(a)} Heaps' law and associated exponents $\beta$ obtained for different values of reinforcement $\delta w$ on a network with $p=0.02$. \textbf{(b)} Exponent $\beta$ as a function of the reinforcement $\delta w$ for networks with different rewiring probabilities $p$.}
	\vspace{-1em}
\end{figure}

\medskip
\textit{Results.} We first test our model on synthetic networks,
and then consider a real case where the underlying
network of relations among concepts can be directly accessed and
used. As a first experiment, based on the idea that concepts are
organized in dense clusters connected by few long-range links, we
model the relations among concepts as a small-world network
(SW)~\cite{watts1999small}. Our choice is supported by recent results
on small-world properties of word associations
\cite{gravino2012complex}, language networks \cite{motter2002topology}
and semantic networks of creative people \cite{benedek2017semantic}.
To construct SW networks we use the procedure proposed in
Ref.~\cite{newman1999scaling}. Namely, we start with a ring of $N$
nodes, each connected to its $2m$ nearest neighbors, and then we add,
with a tunable probability $p$, a new random edge for each of the
edges of the ring. The first thing we want to investigate is the Heaps'
law for the rate at which novelties happen
\cite{heaps1978information,tria2014dynamics}. We therefore 
looked at how the number of distinct nodes $S(t)$ in a sequence
$\mathcal{S}$ generated by a walker grows as a function of length of the
sequence $t$. Figure \ref{fig:sw_model}(a) shows the curves $S(t)$ obtained by
averaging over different realizations of a ERRW process with
reinforcement $\delta w$ on a SW network with rewiring probability
$p=0.02$. All the curves can be well fitted by a power law $S(t) \sim
t^{\beta}$, with an exponent $\beta$ which decreases when the
reinforcement $\delta w$ increases.
Finding the average number of distinct sites visited by a random
walker is a well-known problem in the case of graphs without
reinforcement. In particular, it has been proven that, in the absence of
reinforcement, the average number of distinct sites $S(t)$ visited in
$t$ steps scales
as $S^{\text{ring}}(t) \sim \left(8t/\pi\right)^{{1/2}}$
\cite{dvoretzky19512nd} in one-dimensional lattices and as
$S^{\text{ER}}(t)\sim t$ \cite{de2015average} in 
Erd\H{o}s-R\'{e}nyi random graphs \cite{erdos1959random}. The
transition between these two regimes has been investigated in
Refs.~\cite{jasch2001target,lahtinen2001scaling,almaas2003scaling} for SW networks
with different values of $p$. 
Figure \ref{fig:sw_model}(b) reports the fitted values of the
exponent $\beta$ obtained in the case of ERRW with different strength
of reinforcement.
The four curves refer to SW networks with rewiring probabilities $p=0,
0.02, 0.1$, and $1$. Notice that the previously known results,
$\beta^{\text{ring}}=1/2$ and $\beta^{\text{ER}}=1$, are
recovered as limits of the two curves relative to $p=0$ and $p=1$ when
$\delta w \to 0$. Furthermore, for values of $p$ in the small-world
regime \cite{barrat2000properties}, it is possible to get values of
$\beta$ spanning the entire range $[0,1]$ by tuning the
amount of reinforcement $\delta w$.  This means that the reinforcement
mechanism we propose is able to reproduce all the Heaps' exponents
empirically observed \cite{tria2014dynamics}.

\begin{figure}[t!]
	\vspace{-1em}
	\centering
	\includegraphics[width=0.48\textwidth]
	{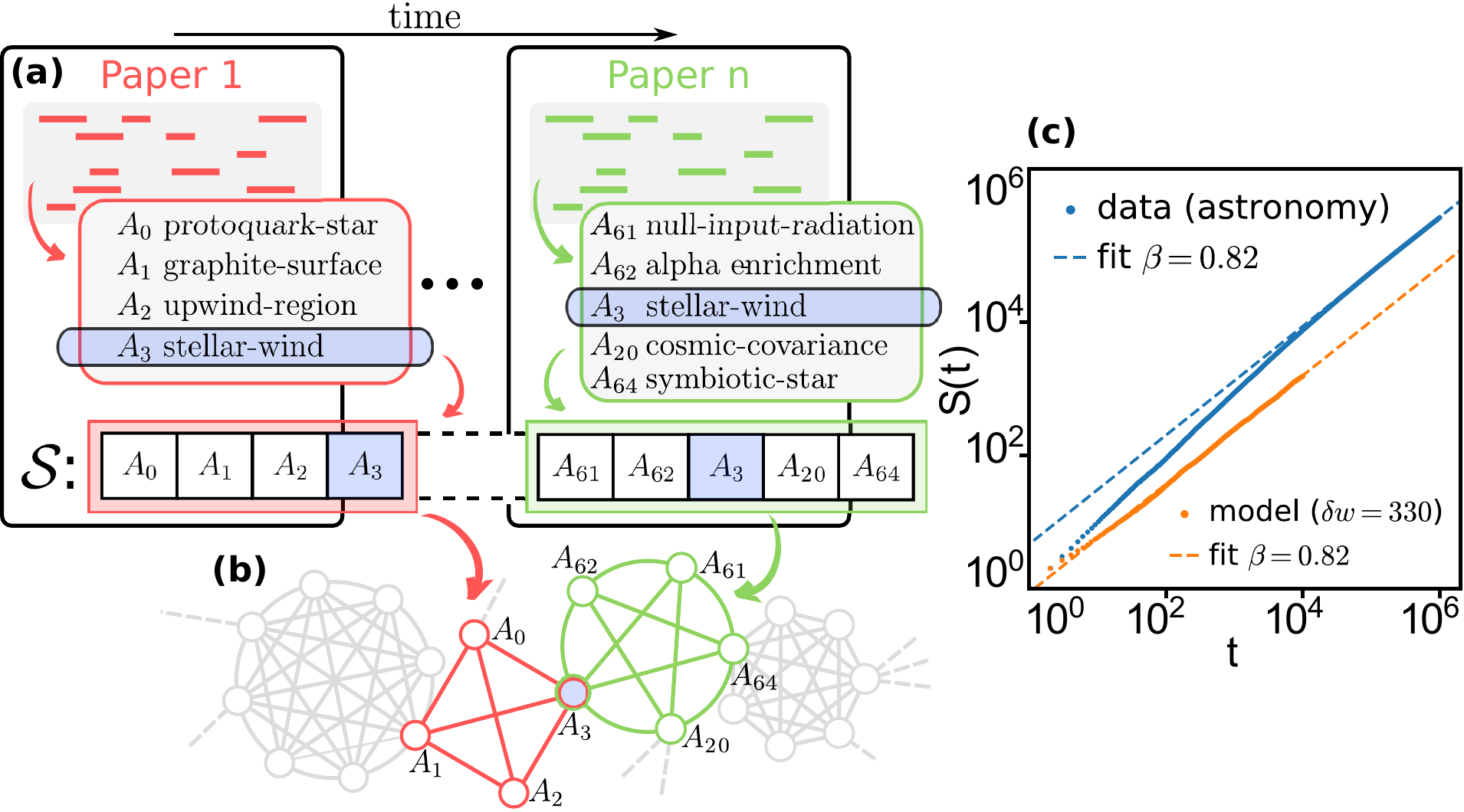}
	\vspace{-1em}
	\caption{\label{fig:empirical_heaps} Growth of knowledge in science. 
			{\bf (a)} For each scientific field, an empirical sequence of scientific concepts $\mathcal{S}$ is extracted from the abstracts of the temporally ordered sequence of papers.
			{\bf (b)} The network of relations among concepts is constructed by linking two concepts if they appear in the same abstract. The network is then used as the underlying structure for the ERRW model.
			{\bf (c)} The model is tuned to the empirical data by choosing the value of the reinforcement $\delta w$ that reproduces the Heaps' exponent $\beta$ associated to $\mathcal{S}$.}
	\vspace{-1em}
\end{figure}

\medskip

\textit{Cognitive growth of science.}  To show how the model works
in a real case, we have extracted the empirical curves $S(t)$
associated with a discovery process on an underlying network whose
topology can be directly accessed.  Specifically, we studied the
growth of knowledge in modern science by analyzing 20 years
(1991-2010) of scientific articles in four different disciplines,
namely, astronomy, ecology, economy and mathematics. Articles were
taken from core journals in these four fields, and bibliographic
records were downloaded from the Web of Science database (details in 
Ref.~\cite{milojevic2012academic}). 
%
%
From a text analysis of each
abstract, we have extracted relevant concepts as
multiword phrases \cite{milojevic2015quantifying} and 
constructed, as in Fig.~\ref{fig:empirical_heaps}(a), 
the real temporal sequence $\mathcal{S}$ in each field from the publication date 
of the papers. Figure \ref{fig:empirical_heaps}(c) shows
that the number $S(t)$ of novel concepts in astronomy grows with the
length $t$ of $\mathcal{S}$ as a power law with a fitted exponent
$\beta=0.82$. Together with the real exploration sequences we have also
extracted, as illustrated in Fig.~\ref{fig:empirical_heaps}(b),
the underlying networks of relations among concepts \cite{baronchelli2013networks} from their
co-occurrences in the abstracts, so that we do not need to rely on 
synthetic small-world topologies, or on the graph version of the UM (see SI of Refs.~\cite{tria2014dynamics,monechi2017waves}).  
Table \ref{table:abs_nets} reports basic 
properties, such as number of nodes $N$, average node degree
$\langle k \rangle$,
characteristic path length $L$ and clustering coefficient
$C$, for the largest components of the four networks we have constructed.
Notice that different disciplines exhibit values of $\langle k \rangle$ ranging
from 19 for mathematics to 172 for astronomy, but all of them have
high values of $C$ and low $L$. We have then run the ERRW on each of
the four networks, tuning the strength of the reinforcement
$\delta w$, the only parameter of the model, so that the obtained curves for the
growth of the number of distinct nodes visited by the walkers
reproduce the empirical values of the exponent $\beta$.
Fig.~\ref{fig:empirical_heaps}(c) shows that, for the case of
astronomy, the curve $S(t)$ of our model with $\delta w = 330$ has a
power-law growth with exponent $\beta= 0.82$, equal to the one
extracted from the real sequence of concepts. The values of reinforcement
obtained for the other scientific disciplines are reported in
Table \ref{table:abs_nets}.
Notice that stronger reinforcement is required to get the same $\beta$ in networks with higher values of $\langle k \rangle$  (see \cite{SuppMat}).
\definecolor{LightBlue}{rgb}{0.78,0.84,0.98}
\newcolumntype{b}{ >{\columncolor{LightBlue}[0.95\tabcolsep]} c } 
{
	\setlength\extrarowheight{2pt}
	\begin{table} 
		\begin{center} 
			\begin{tabular}{|c|c|c|c|c|c|c|b|} 
				\hline
				{\bf Research field} & Papers & $N$& $\langle k\rangle$ & $C$ & $L$ & $\beta$ & $\delta w$ \\
				\hline
				Astronomy &97,255 &103,069 &172 &0.41 &2.48 &0.82 &330 \\
				\hline
				Ecology &18,272 &289,061 &52 &0.89 &2.98 &0.85 &105 \\
				\hline
				Economy &7,100 &60,327 &20 &0.91 &3.69 &0.91 &6 \\
				\hline
				Mathematics &7,874 &48,593 &19 &0.89 &3.69 &0.87 &20  \\
				\hline
			\end{tabular}
			\caption[]{Statistics of the network of concepts in four research fields, together with the empirical Heaps' exponent $\beta$ and the value of $\delta w$ that reproduces it.}
			\vspace{-1em}
			\label{table:abs_nets}
		\end{center}
		\vspace{-1em}
	\end{table}
}

\begin{figure}[t]
	\vspace{-1em}
	\centering
	\includegraphics[width=0.45\textwidth]{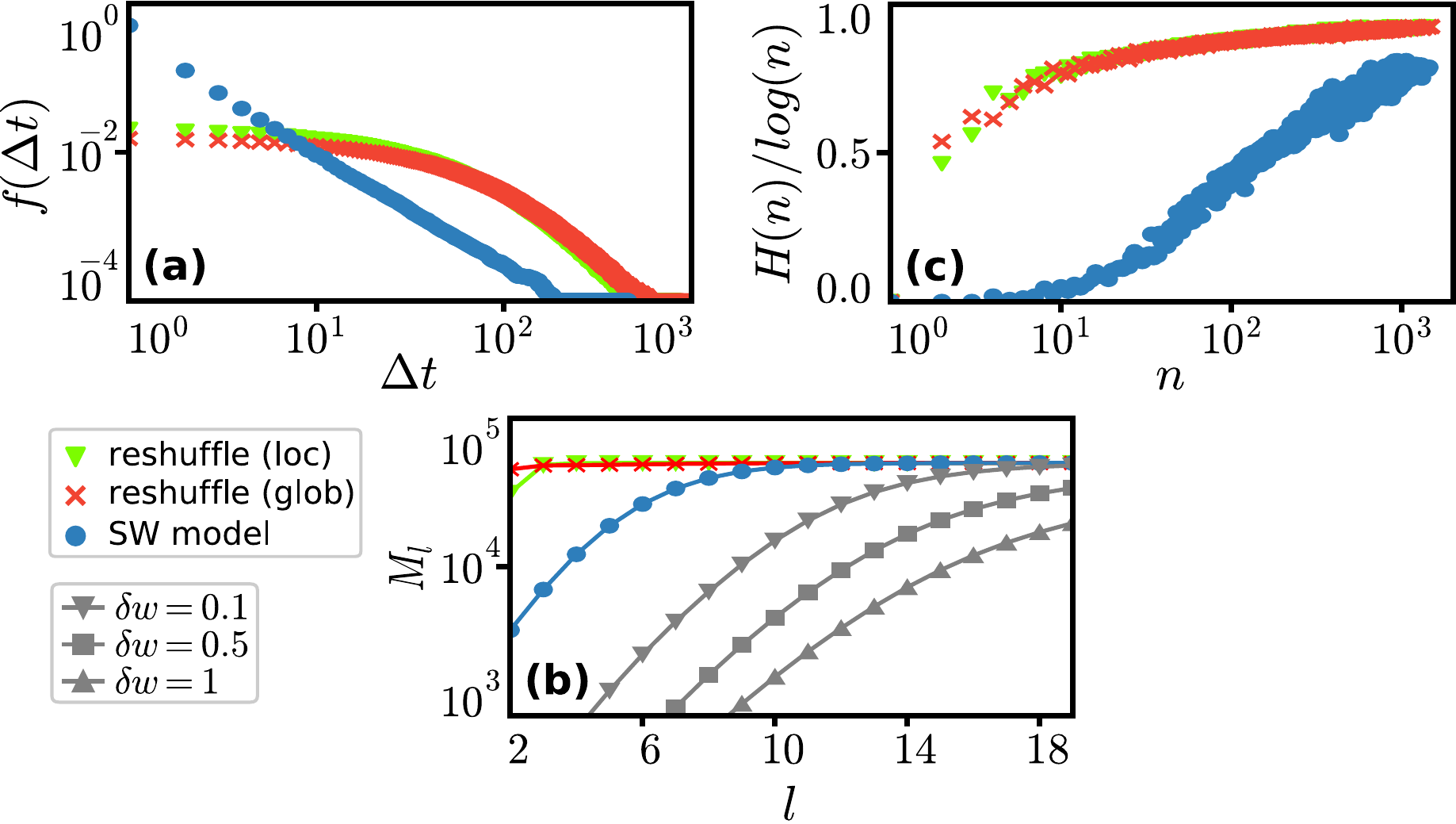}
	\vspace{-1em}
	\caption{\label{fig:correlations} Correlations among concepts
		produced by an ERRW ($\delta w = 0.01$) on a SW network
		($p=0.02$). \textbf{(a)} Frequency distribution of
		interevent times $\Delta t$ between consecutive occurrences
		of the same concept (node in our model). \textbf{(b)} Number
		$M_{l}$ of different subsequences of length $l$ as a
		function of $l$. \textbf{(c)} Normalized entropy of the
		sequence of visited nodes as a function of $n$, the number
		of times the nodes have been visited. In each panel, blue circles show
			average values over 20 different realizations, while triangles and crosses refer 
			to those of (globally and locally) reshuffled sequences.}
	\vspace{-1em}
\end{figure}

\medskip
\textit{Correlations.}  In addition to the Heaps' law, our model
naturally captures also the correlations among novelties, which are a hallmark of real
exploration sequences \cite{tria2014dynamics,monechi2017waves}.
The results for sequences generated by the ERRW model on SW
	networks with $p=0.02$ and $\delta w = 0.01$ are plotted in Figure
	\ref{fig:correlations} (different values of $p$ and $\delta w$ in Supp.~Mat. \cite{SuppMat}).  In particular, Fig.~\ref{fig:correlations}(a) shows that the frequency distribution
$f(\Delta t)$ of interevent times $\Delta t$ between pairs of
consecutive occurrences of the same concept is a power law, like the
ones found for novelties in Wikipedia and in other data sets in
Refs.~\cite{tria2014dynamics,monechi2017waves}.  Furthermore, the
	shape of $f(\Delta t)$ in our model significantly differs from that
	obtained by reshuffling the sequences locally and globally (see \cite{SuppMat}).  Notice that $f(\Delta t)$ is the distribution of first
	return times (FRT), and it remains an interesting research question
	to investigate how FRT are linked to first passage times (FPT) in
	the case of correlated random walks.

\noindent We have also looked at how $M_{l}$, the number of distinct subsequences
	of $\mathcal{S}$ of length $l$, grows with $l$ \cite{ebeling1992word}. 
In Fig.~\ref{fig:correlations}(b) the curve $M_{l}$ generated by the
ERRW model with $\delta w = 0.01$ is compared to those obtained by
reshuffling the sequences. The value of $M_{l}$ grows with $l$, until
it reaches a plateau (equal to $T-l$, where $T=5\times 10^4$ is the
number of steps of the walker in the simulation) as a consequence of
the finite length of $\mathcal{S}$. Interestingly, the analogous
curves for the null models immediately approach the saturation value,
meaning that a process without reinforcement would generate all the
possible subsequences in a sequence of length $T$, while with the
reinforcement this number drops down because of the correlations.
In our model, the correlated sequences naturally emerge
	from the co-evolution of network and walker dynamics, 
	while the UM \cite{tria2014dynamics} requires the introduction of an
	additional semantic triggering mechanism to reproduce the correlations
	found in the data (see Supp.~Mat.~\cite{SuppMat} for a detailed discussion
	of the differences between the two models).
To better characterize the correlations, we studied how homogeneously
concepts occur in the sequence $\mathcal{S}$, after their first
appearance. Following Tria {\em et al.}~\cite{tria2014dynamics}, we
have divided the sequence $\mathcal{S}$ in $n^{(A)}$ subsequences of
the same length, with $n^{(A)}$ being the total number of occurrence
of $A$ in $\mathcal{S}$, and we have evaluated
the Shannon entropy \cite{shannon2001mathematical}
	$H^{(A)}=-\sum_{s=1}^{n^{(A)}} p^{(A)}_{s}\log p^{(A)}_{s}$ for every concept $A$,
where $p^{(A)}_{s}=n^{(A)}_{s}/n^{(A)}$ denotes the probability of
finding concept $A$ in subsequence $s$. Figure \ref{fig:correlations}(c) shows the
normalized average entropy $H(n)$ of concepts appearing $n$
times. 
Again, the large differences with respect to the null models 
reveal the correlated dynamics of our model.
Similar results are obtained 
	for the network of relationships among scientific concepts \cite{SuppMat}, 
	confirming the validity of the choice of SW networks as underlying structures.

\medskip

In summary, the mechanism of coevolution of network and random walks 
introduced in this work naturally reproduces all the properties
observed in real innovation processes, including the correlated nature
of exploration trajectories. With the topology of the network being a key
ingredient of the model, we hope our framework will be found 
useful in all cases where the network
can be directly reconstructed from data, as in the
study of scientific innovations reported here. 

\bigskip

\begin{acknowledgments}
	We acknowledge support from EPSRC Grant No. EP/N013492/1.
\end{acknowledgments}

\clearpage

\widetext
\begin{center}
	\textbf{\large Supplemental material: Network dynamics of innovation processes}
\end{center}
\setcounter{figure}{0}
\setcounter{table}{0}
\makeatletter
\renewcommand{\thefigure}{S\arabic{figure}}

\section{NULL MODELS: reshuffling the sequences}\label{sec:null}

In the main text, in order to check whether the sequences produced by
our ERRW model are correlated, we have compare them to reshuffled
versions of the sequences.  More precisely, given a trajectory
$\mathcal{S}$ of visited nodes (concepts), it is possible to define
two null models based on the following two reshuffling procedures
\cite{tria2014dynamics}. The simplest procedure consists in the global
reshuffling of all the elements of $\mathcal{S}$ (indicated as
\textit{glob} in Figure 4 of the main text). This method destroys
indeed the correlations (if there are any) in the sequence, but it also 
modifies the rate at which the new concepts appear, ultimately
changing the exponent of the Heaps' law. Contrarily,
the rate can be preserved by
defining a second version of the null model, based on a local
reshuffling (indicated as \textit{loc} in Figure 4 of the main
text). In this second procedure we reshuffle all the elements
in $\mathcal{S}$ only after their
first appearance, such that a concept cannot be randomly replaced in
the sequence before the actual time it has been discovered.

\section{CORRELATIONS produced by ERRWs on real networks}\label{sec:corr_real}

\begin{figure*}
	\includegraphics[width=1\textwidth]{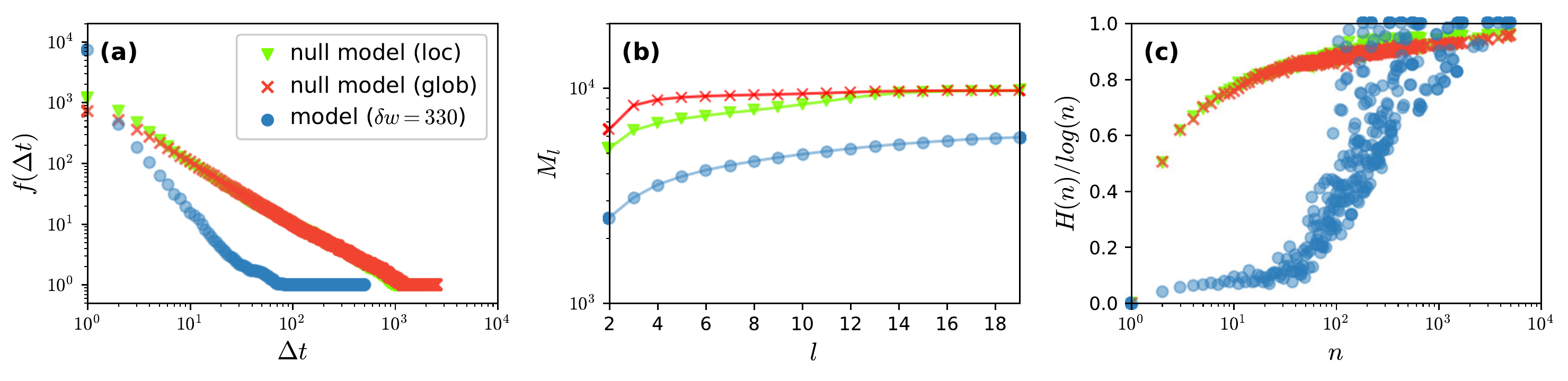}
	
	\caption{\label{SI_abs_corr} Correlations among concepts for
		the growth of knowledge in science (Astronomy shown)
		produced by an ERRW model. The ERRW is tuned to the
		empirical data by selecting the reinforcement $\delta w$
		that reproduces the Heaps' exponent $\beta$ obtained by
		fitting the associated Heaps' curve as a power law (for the
		Astronomy case shown $\delta w=330$).~\textbf{(a)} Frequency
		distribution of inter-event times $\Delta t$ between
		consecutive occurrences of the same concept (node in our
		model).~\textbf{(b)} Number $M_{l}$ of different
		subsequences of length $l$ as a function of
		$l$.~\textbf{(c)} Normalized entropy of the sequence of
		visited nodes as a function of $n$, the number of times the
		nodes have been visited (see the main text for details). In
		each panel, blue circles show average values over 20
		different realizations, while triangles and crosses refer to those of (globally and locally)
		reshuffled sequences.}
\end{figure*}

In the main text, we have shown how the ERRW model on small-world (SW)
networks is
able to produce correlated sequences of concepts.  We have also
proposed a study case of the ERRW model on real topologies 
extracted from empirical data. In particular, we have explored the
cognitive growth of science by extracting empirical sequences of
relevant concepts in different scientific fields. 
For each of the fields considered, we have then tuned the
reinforcement parameter of our model 
in order to
produce sequences with the same Heaps' exponents as the empirical ones
(see Figure 3 and Table 1 of the main text). Here, we investigate 
correlations in the sequences produced by ERRWs on real networks.
Figure \ref{SI_abs_corr} reports the same quantities we used 
to study correlations in sequences produced by ERRW on synthetic 
small-world networks (see Figure 4 of the main text), namely the average
entropy of the sequence (Figure \ref{SI_abs_corr}(a)), number $M_l$ of
different subsequences of length $l$ as a function of $l$ (Figure
\ref{SI_abs_corr}(b)), and frequency distribution $f(\Delta t)$ of
inter-event times $\Delta t$ between couples of consecutive
concepts (Figure \ref{SI_abs_corr}(c)). In each plot, results are
compared to the two null models defined in Section \ref{sec:null} of
this Supplemental Material, confirming the correlated nature of the
sequences. Furthermore, the comparison with the same statistics
obtained for ERRWs on SW networks (see Figure 4 of the main text)
confirms again that small-world topologies represent a good choice
for modeling the relations among concepts.

\section{CORRELATIONS produced by ERRWs on synthetic networks}

\begin{figure*}
	\includegraphics[width=1\textwidth]{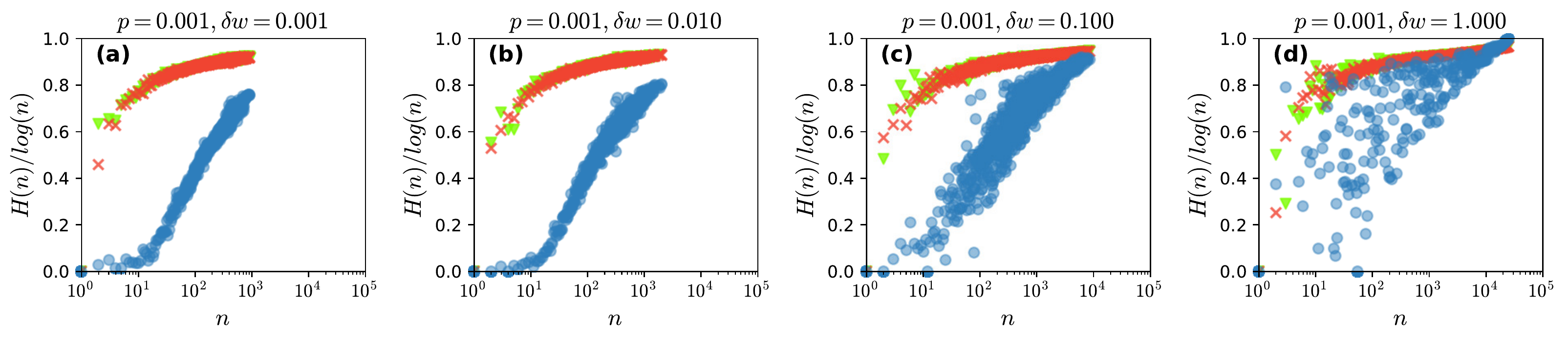}
	\includegraphics[width=1\textwidth]{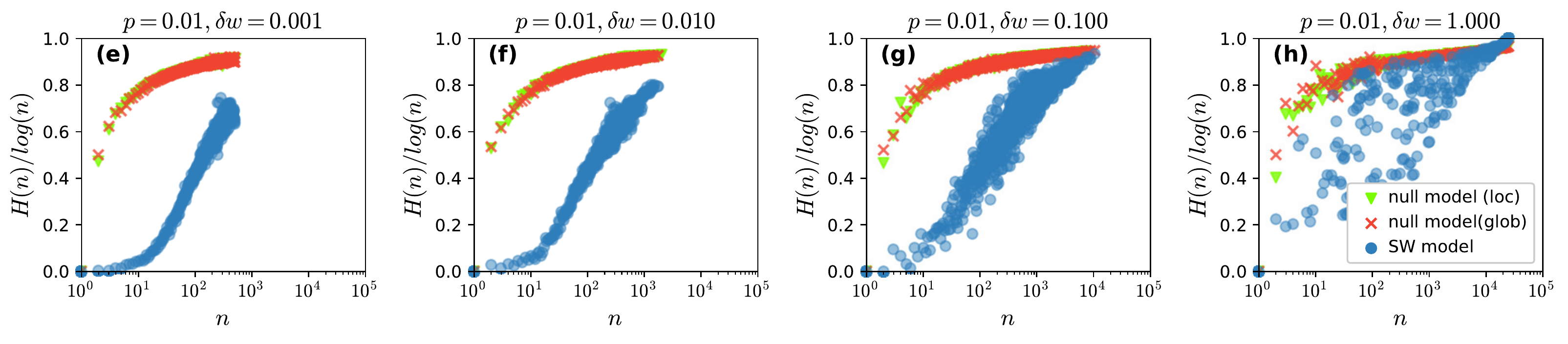}
	\includegraphics[width=1\textwidth]{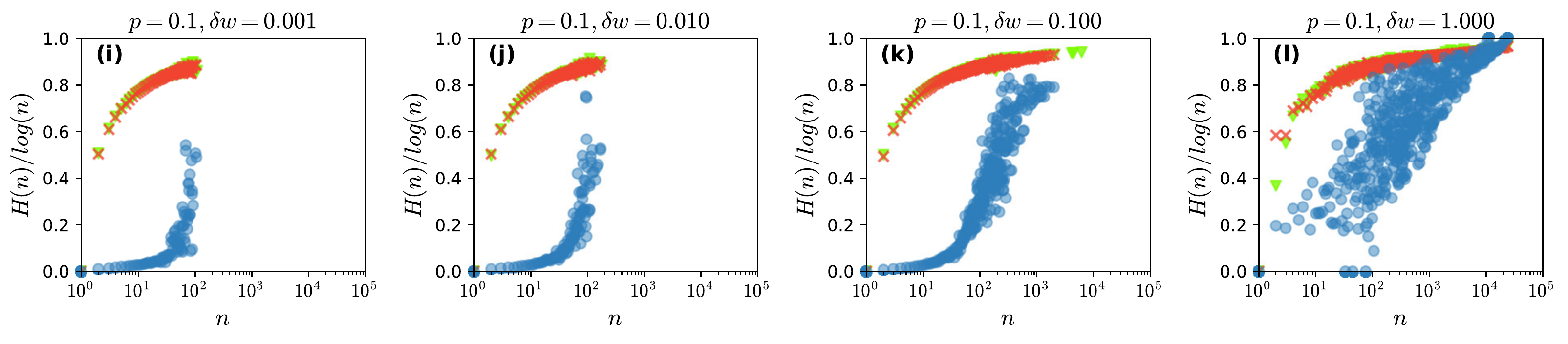}
	
	\caption{\label{SI_en} Correlations among concepts produced by an edge-reinforced random walk on a SW network for different values of link probability $p$ and reinforcement $\delta w$ (see the main text for details). Normalized entropy of the sequence of visited nodes as a function of $n$, the number of times the nodes have been visited. In each panel, blue circles show average values over 20 different realizations, while triangles and crosses refer to those of (globally and locally) reshuffled sequences.}
\end{figure*}

\begin{figure*}
	\includegraphics[width=1\textwidth]{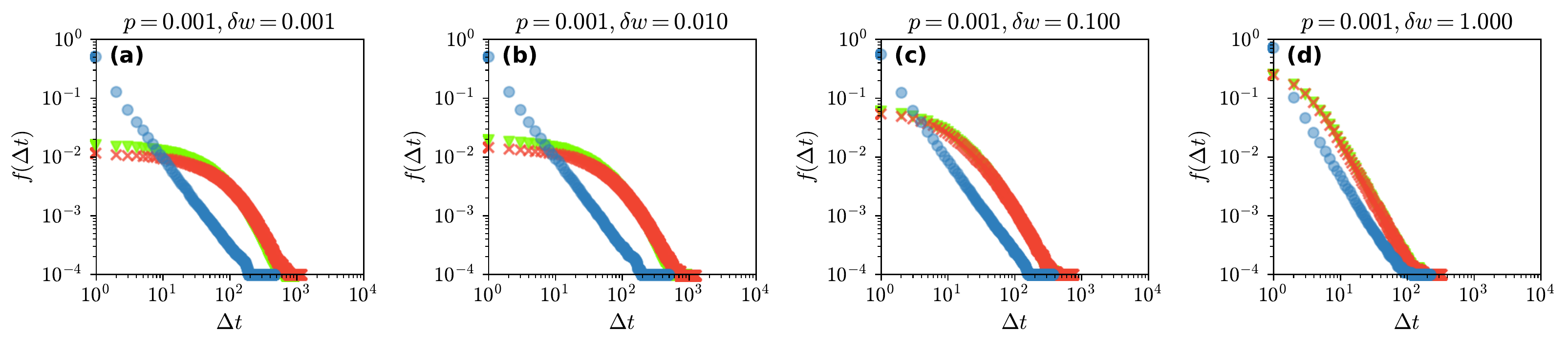}
	\includegraphics[width=1\textwidth]{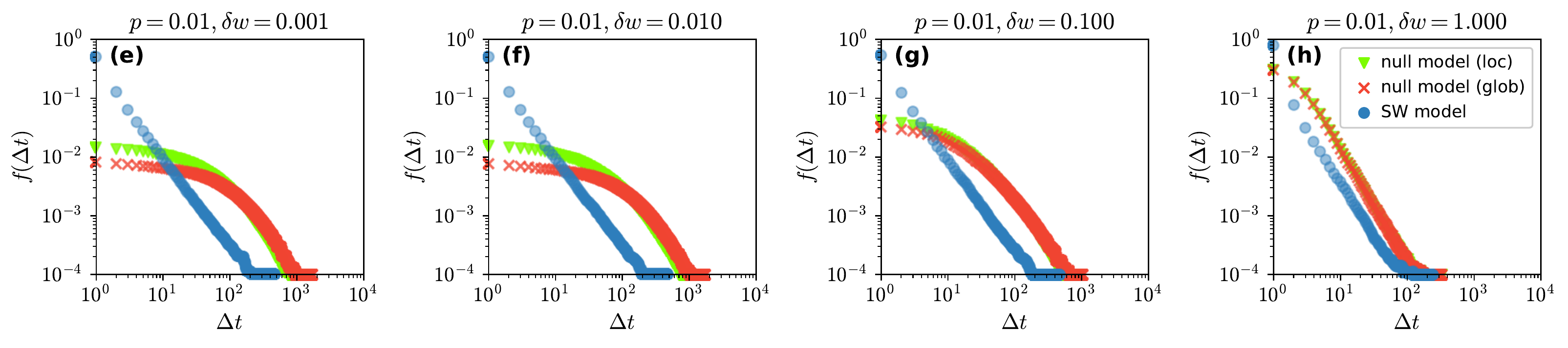}
	\includegraphics[width=1\textwidth]{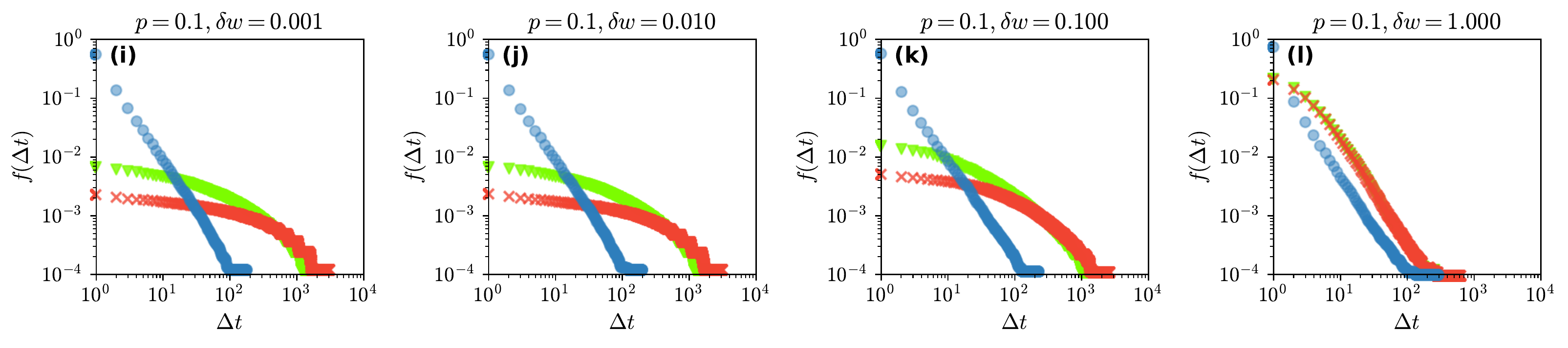}
	
	\caption{\label{SI_int} Correlations among concepts produced
		by an edge-reinforced random walk on a SW network for
		different values of link probability $p$ and reinforcement
		$\delta w$ (see the main text for details). Frequency
		distribution of inter-event times $\Delta t$ between
		consecutive occurrences of the same concept (node in our
		model). In each panel, blue circles show average values over 20 different realizations,
		while triangles and crosses refer to those of (globally and locally) reshuffled sequences.}
\end{figure*}

In the main text we have implemented the ERRW model on small-world
networks, which proved to be good topologies for modeling the structure
of relations among concepts (see Section \ref{sec:corr_real} of this
text and
Refs~\cite{gravino2012complex,motter2002topology,benedek2017semantic}). In
addition to the plots in Figure 4 of the main text, where we studied
the correlations produced by an edge-reinforced random walk over a SW
network with fixed link probability $p$ for a fixed amount of
reinforcement at $\delta w=0.01$, here we show the curves of average
entropy of sequence (Figure \ref{SI_en}) and frequency distribution
$f(\Delta t)$ of inter-event times $\Delta t$ between couples of
consecutive concepts (Figure \ref{SI_int}) for different values of
reinforcement, ranging from $\delta w=0.001$ to $\delta w=1$. Three
different cases of SW networks with $N=10^6$ nodes and respectively
with link rewiring 
probability $p=0.001$ (Fig.~\ref{SI_en}(a-d) and
Fig.~\ref{SI_int}(a-d)), $p=0.01$ (Fig.~\ref{SI_en}(e-h) and
Fig.~\ref{SI_int}(e-h)) and $p=0.1$ (Fig.~\ref{SI_en}(i-l) and
Fig.~\ref{SI_int}(i-l)), are considered.
All the curves are compared to the corresponding 
null models as defined in Section \ref{sec:null} of this Supplemental
Material.
\bigskip

\section{The effect of the average degree on the reinforcement}

\begin{figure*}
	\includegraphics[width=0.6\textwidth]{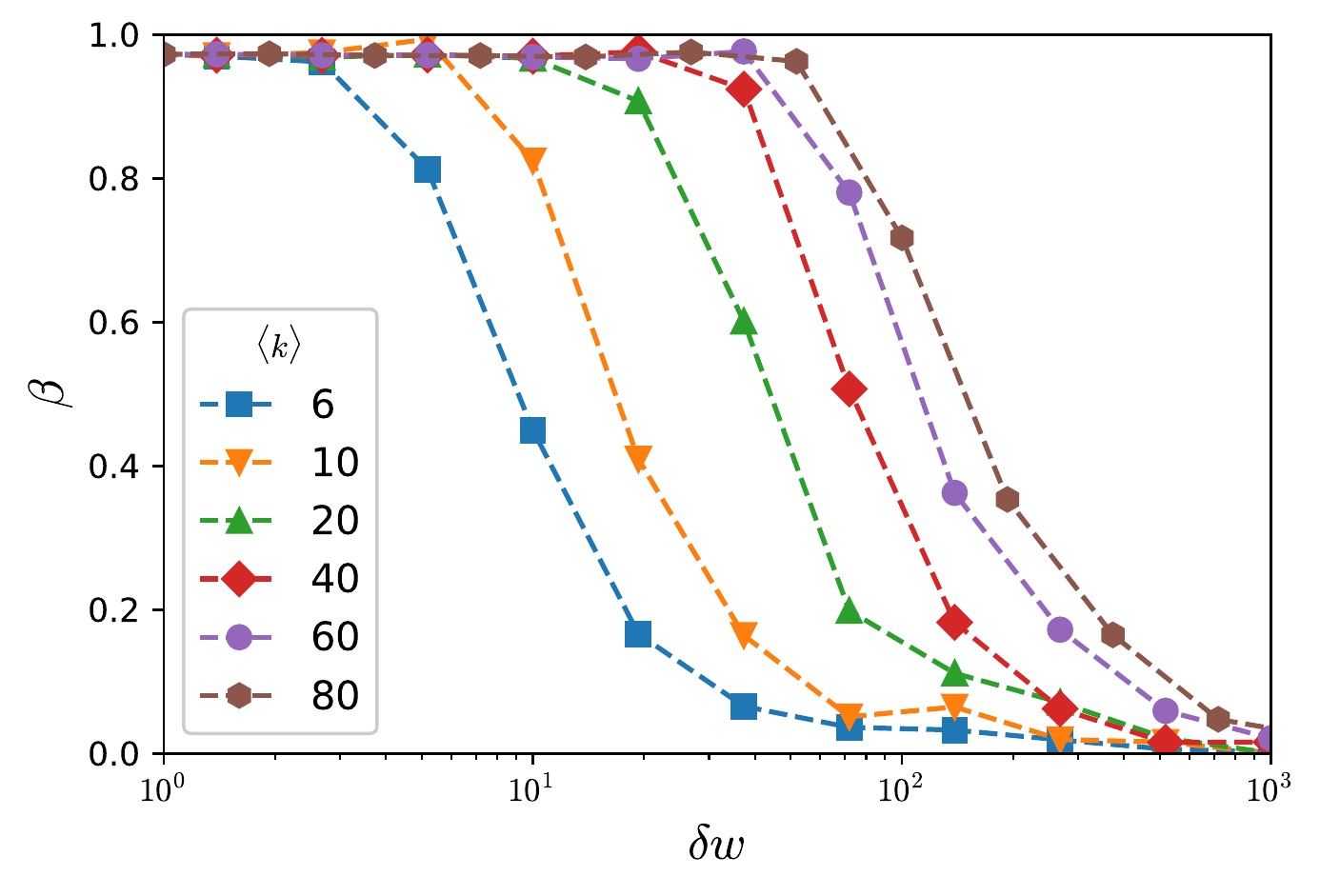}
	\caption{\label{SI_avg_k} ERRW on ER networks with $N=10^5$ and average degree $\langle k \rangle$. Heaps' exponent $\beta$ as a function of reinforcement $\delta w$.
	}
\end{figure*}

To better understand the wide range of values obtained for the
reinforcement parameter from the analysis of the
growth of knowledge in different scientific
fields (see Table 1 of the main text), we looked at the relation 
between the exponent $\beta$ extracted from the Heaps' law 
and the reinforcement $\delta w$ in networks
with different average node degree.
Figure \ref{SI_avg_k} shows $\delta w$ 
versus $\beta$. Each curve corresponds to Erd\H{o}s-R\'enyi
random graphs with $N=10^5$ nodes and 
average degrees $\langle k \rangle$ ranging from 6 to 
80. As expected, the
average degree significantly impacts the reinforcement. In particular,
the higher the value of $\langle k \rangle$, the stronger the reinforcement
$\delta w$ has to be in order to produce the same Heaps' exponent.
This is easily understandable if one considers the possible 
choices of a walker reaching a node connected to a
link that has been reinforced. If the node has a
high degree, the probability of selecting that specific link among all
the others will be smaller, and the walker will more easily select a
new link, leading to a previously undiscovered node, and therefore to
a higher $\beta$. If one wants to keep a certain discovery rate in 
networks with higher $\langle k \rangle$, higher values of
reinforcement will then need to be considered.

\section{Comparing ERRWs to the network version of the urn models}

Here we clarify some aspects regarding  similarities and differences
between our ERRW model and the urn models proposed by Tria
et al. \cite{tria2014dynamics}, together with their network versions.
\\
In the main text, we state that for the edge-reinforced random walk
(ERRW) model, the conditional probability
$\text{Prob}\left[X_{t+1}=i|i_{0},i_{1},\ldots,i_{t}\right] $ that, at
time step $t+1$, the walker is at node $i$, after a trajectory
$\mathcal{S}=(i_{0},i_{1},i_{2},\ldots,i_t)$, depends on the whole
history of the visited nodes, namely on the frequency but also on the
precise order in which they have been visited. This is different
from what happens in the basic version of the urn model.
Using the notation introduced by Tria et al. \cite{tria2014dynamics},
in the main text, by urn model (UM) we referred to the basic urn model,
i.e. the urn model without semantic. In this case, each ball in the
urn has the same probability of being extracted. Since there might be
multiple balls of the same color, the probability to extract a given
color will depend on the number of balls of that color, and also on
the total number of balls in the urn. The number of balls of a given
color at time $t$ depends on how many times balls of that color have
been extracted up to time $t$ (i.e. on how many times the color has
been reinforced), but it does not depend on the specific order of
appearance in the sequence of extracted balls. The number of balls in
the urn at time $t$ depends on the number of balls initially present
in the urn, plus the ones added by mean of the reinforcement mechanism
($\rho$ additional balls for every $t$), plus the balls representing
the ``adjacent possible" ($\nu +1$ additional balls, every time a
color is extracted for the first time).\\

For example, let us consider the UM with parameters $\rho=1$ and $\nu=0$, and let us 
indicate as $R$, $B$, $G$ balls respectively of color Red, Blue and Green. By $\mathcal{U}_t$ we indicate the urn at time $t$, while $\mathcal{S}_t$ represents the sequence of extracted colors from the urn at time $t$, which will trigger a reinforcement at $t+1$ of $\rho=1$ new balls of color $X$ every time a ball of color $X$ is extracted, and a 
further addition of $\nu+1 =1$ balls of new colors every time a color is extracted for the first 
time (novelty).\\

A possible evolution, starting from an initial condition with one red ball in the urn at time $t=1$, is the following: 

\noindent
At $t=1$, $\mathcal{U}_1= \{  R \}$. A  $R$ ball is drawn: $\mathcal{S}_1= (R)$. $R$ is reinforced and $B$ is added to the urn. \\
At $t=2$, $\mathcal{U}_2= \{ R,R,B \}$. A $B$ ball is drawn: $\mathcal{S}_2= (R,B)$. $B$ is reinforced and $G$ is added;\\
At $t=3$, $\mathcal{U}_3= \{ R,R,B,B,G \}$. A $R$ ball is drawn: $\mathcal{S}_3 = (R, B, R)$. $R$ is reinforced;\\
At $t=4$, $\mathcal{U}_4= \{ R,R,R,B,B,G \}$. A $R$ ball is drawn: $\mathcal{S}_4= (R, B, R, R)$. $R$ is reinforced;\\
At $t=5$, $\mathcal{U}_5= \{ R,R,R,R,B,B,G \}$. A $B$ ball is drawn: $\mathcal{S}_5=(R, B, R, R, B)$. $B$ is 
reinforced;\\
At $t=6$, $\mathcal{U}_6= \{ R,R,R,R,B,B,B,G \}$. 

\noindent
Now, the probabilities of extracting balls of different colors at time $t=6$ are respectively: $p_{R}=1/2, p_{B}=3/8$ and $p_{G}=1/8$.\\

Notice that another possible evolution, starting from the same initial condition, is the following: 

\noindent
At $t=1$, $\hat{\mathcal{U}}_1= \{  R \}$. A  $R$ ball is drawn: $\hat{\mathcal{S}}_1= (R)$. $R$ is reinforced and $B$ is added to the urn. \\
At $t=2$, $\hat{\mathcal{U}}_2= \{ R,R,B \}$. A $R$ ball is drawn: $\hat{\mathcal{S}}_2= (R,R)$. $R$ is reinforced;\\
At $t=3$, $\hat{\mathcal{U}}_3= \{ R,R,R,B \}$. A $B$ ball is drawn: $\hat{\mathcal{S}}_3 = (R, R, B)$. $B$ is reinforced and $G$ is added;\\
At $t=4$, $\hat{\mathcal{U}}_4= \{ R,R,R,B,B,G \}$. A $B$ ball is drawn: $\hat{\mathcal{S}}_4= (R, R, B, B)$. $B$ is reinforced;\\
At $t=5$, $\hat{\mathcal{U}}_5= \{ R,R,R,B,B,B,G \}$. A $R$ ball is drawn: $\hat{\mathcal{S}}_5=(R, R, B, B, R)$. $R$ is reinforced;\\
At $t=6$, $\hat{\mathcal{U}}_6= \{ R,R,R,R,B,B,B,G \}$. 

\noindent
Although the two sequences generated at time $t=5$ are different, namely $\hat{\mathcal{S}}_5 \neq \mathcal{S}_5 $, they contain the same number of entries for each color, and the two urns at time $t=6$ will be equal, namely $\hat{\mathcal{U}}_6=\mathcal{U}_6$, so that the probabilities of extracting balls of different colors at time $t=6$ will be $p_{R}=1/2, p_{B}=3/8$ and $p_{G}=1/8$ also for the second urn evolution. 

With this simple example we have been able to show that the probability of extracting a color at a given time depends on the number of balls of each color, but not on the precise order of the extracted balls. \\

Our focus until now has been on the basic UM proposed by Tria et al.
There is however a more refined version of the model proposed in
Ref.~\cite{tria2014dynamics}, called urn model with {\it semantic
	triggering}, from now on UMS. In this second version, the authors
propose an urn model that is also able to reproduce the correlations 
of empirical sequences. The model is based on the introduction of  
{\it semantic labels} attached to the balls (different balls and 
colors might share the same label), together with a mechanism
named semantic triggering. The semantic triggering mechanism
is able to produce correlated sequences, but it also requires the
addition of a third parameter, namely $\eta$, to the model. Notice,
instead, that the model we propose in this paper does not need labels
or additional mechanisms. In our model correlations emerge naturally
from the co-evolution of the walker dynamics and the network.

Finally, in the Supplementary Information of
Ref.~\cite{tria2014dynamics} the authors discuss how to map urn models
into a growing network framework. Such a 
mapping is exact only in the case when $\eta=1$, which actually
corresponds to the simple UM without semantic and thus without
correlations.  Contrarily, when $\eta \le 1$, i.e. in the case of the
UMS in which the model is able to produce correlated sequences,
the mapping is not one-to-one. The key difference is in fact
that in a network the
connections are always well defined (a link exists or not). In fact,
the possibility of going from a node $n_{A}$ to any other node is
restricted to the neighbors of $n_A$, while for the case of the urn
model the possibility of drawing any ball $X$ after the extraction of
a given ball $A$ is always probabilistic. As a consequence, the
network framework of the urn model presented in S.I. of
Ref.~\cite{tria2014dynamics} works exactly only for the very specific
case $\eta=1$ corresponding to a fully connected
network (where a walker can move from each node to every other node,
in the same way as any ball can be drawn from an urn after the
extraction of any other ball).

\end{document}